\documentclass[preprint,12pt]{elsarticle}

\usepackage{amssymb}

\usepackage{lineno}
\usepackage[separate-uncertainty=true]{siunitx}
\usepackage{float}
\DeclareSIUnit\GeV{GeV}
\DeclareSIUnit\c{c}
\DeclareSIUnit\QDCunits{QDC units}
\DeclareSIUnit\qdc{\text{QDC-U}}
\DeclareSIUnit\photon{photon}
\DeclareSIUnit\bit{bit}
\usepackage{hyperref}

\begin{document}
\begin{frontmatter}



\title{Cherenkov Diffraction Radiation Emissions from Single Electrons and Positrons on a Fused Silica Radiator}

\author[wvs]{Silas Ruhrberg Estevez}
\author[wvs]{Tobias Baumgartner}
\author[wvs]{Johann Bahl}
\author[wvs]{Thomas Lehrach}
\author[wvs]{Tobias Thole}
\author[wvs]{Benildur Nickel}
\author[wvs]{Philipp Loewe}
\author[wvs]{Lukas Hildebrandt}
\author[cern]{Cristóv\~ao Beir\~ao da Cruz E Silva}
\author[desy]{Paul Schütze}
\author[cern]{Markus Joos\corref{cor1}}

\cortext[cor1]{Corresponding author:   markus.joos@cern.ch (Markus Joos) }

\affiliation[wvs]{organization={Werner-von-Siemens-Gymnasium},
            addressline={Beskidenstraße 1}, 
            city={Berlin},
            postcode={14129}, 
            country={Germany}}

\affiliation[desy]{organization={DESY},
            addressline={Notkestraße 85}, 
            city={Hamburg},
            postcode={22607}, 
            country={Germany}}

\affiliation[cern]{organization={CERN},
            addressline={Esplanade des Particules 1},
            city={Meyrin},
            postcode={1217},
            state={Geneva},
            country={Switzerland}}

\begin{abstract}
Beam diagnostics are crucial for smooth accelerator operations. Many techniques rely on instrumentation in which the beam properties are significantly affected by the measurement. Novel approaches aim to use Cherenkov Diffraction Radiation (ChDR) for non-invasive diagnostics. Unlike regular Cherenkov Radiation, the charged particles do not have to move inside of the medium, but it is sufficient for them to move in its vicinity as long as they are faster than the speed of light in the medium. Changes to the beam properties due to ChDR measurements are consequently negligible. To examine ChDR emission under different conditions, we placed a fused silica radiator in the DESY II Test Beam. We observed a linear increase in ChDR intensity for electron and positron momenta between $\SI{1}{\GeV \per \c}$ and $\SI{5}{\GeV \per \c}$. Additionally, we found that electrons produce significantly more ChDR than positrons for increasing particle momenta. The results suggest a need for further research into the ChDR generation by electrons and positrons and  may find application in the design of future beam diagnostic devices.
\end{abstract}

\end{frontmatter}


\section{Introduction}
 Cherenkov radiation is light produced by charged particles when they pass through an optically transparent medium at speeds exceeding the speed of light in that medium~\cite{CRdefinition}. It was first observed experimentally in 1937 by Pavel Cherenkov~\cite{Cherenkov1937}. He shared the Nobel Prize in Physics 1958 with Ilya Frank and Igor Tamm who developed a theoretical model of the phenomenon~\cite{Tamm1939}. The model was improved by Ginzburg and Frank to show the emission originated from dielectric material regions parallel to the particle motion \cite{Ginzburg1947}. The radiation emisison has since been calculated using electromagnetic field eigenvalues ~\cite{Linhart1955} and Di Francia expansions ~\cite{Ulrich1966}. More recent work has described ChDR generation in different scenarios~\cite{Harryman2020}\cite{Lasocha2020}.\\
 
 In the last few years, the existence of ChDR has been proven experimentally~\cite{Kieffer2018}. ChDR can be emitted if an ultrarelativistic charged particle moves in the vicinity of a dielectric medium~\cite{Alves2019}. The atoms of the medium get polarized by the electric field of the ultrarelativistic charged particle, oscillate, and thereby emit light~\cite{Bobb2018} at a characteristic Cherenkov angle $\cos(\theta)=\frac{1}{\beta n}$, where $\beta$ is the relativistic factor and $n$ is the refractive index of the material. For fused silica $(n=1.46)$ and ultrarelativistic particles $(\beta \approx 1)$ the angle is approximately $\SI{46.8}{\degree}$~\cite{Lefevre2018}.\\
 
ChDR has been proposed to be a method for non-invasive beam diagnostics as the particles do not physically interact with the radiator~\cite{Kieffer2020}. Beam position and bunch length monitors exploiting ChDR emission have been trialled successfully~\cite{Alves2019}\cite{Curcio2020}. In this article we present the results of placing a dielectric radiator in the vicinity of a particle beam at the DESY II Test Beam Facility and measuring the emission rates of photons under different conditions\footnote{All experiments were conducted by high school students under expert guidance as part of the Beamline for Schools (BL4S) competition 2020. BL4S is a worldwide competition offered by CERN since 2014 that provides high school students with the opportunity to conduct their own experiments at a state-of-the-art particle accelerator~\cite{Arce-Larreta2021}.In the years 2019 - 2021, BL4S was co-organized by DESY and held mostly at their facilities in Hamburg due to the Long Shutdown 2 at CERN~\cite{Aretz2020}.}. We focus on a comparison of the emissions from electrons and positrons in the same setup. To our knowledge this has not been done before as previous experiments were conducted on circular colliders where electrons and positrons travel in opposite directions~\cite{Kieffer2020}.

\section{Methods}
\subsection{Experimental setup}
The DESY II Test Beam Facility offers positron and electron beams with selectable momenta from $\SI{1}{\GeV\per\c}$ to $\SI{6}{\GeV\per\c}$~\cite{Diener2018}. A maximum particle rate of $\SI{10d3}{\Hz}$ is reached at around $\SI{2}{\GeV\per\c}$~\cite{Aretz2020}. The Test Beam is generated by double conversion of the DESY II synchrotron beam~\cite{Diener2018}. Bremsstrahlung is produced from $\SI{7}{\micro\metre}$ carbon primary targets held inside the synchrotron beam. The Bremsstrahlung then creates electron positrons pairs on a secondary metal target. The particles subsequently pass through a dipole magnet, which allows selection of particle type and momentum.  A $\SI{10}{\mm} \times \SI{20}{\mm}$ collimator narrows the beam before it traverses the experimental setup.\\

\begin{figure}
\centering
 \includegraphics[width=0.5\textwidth]{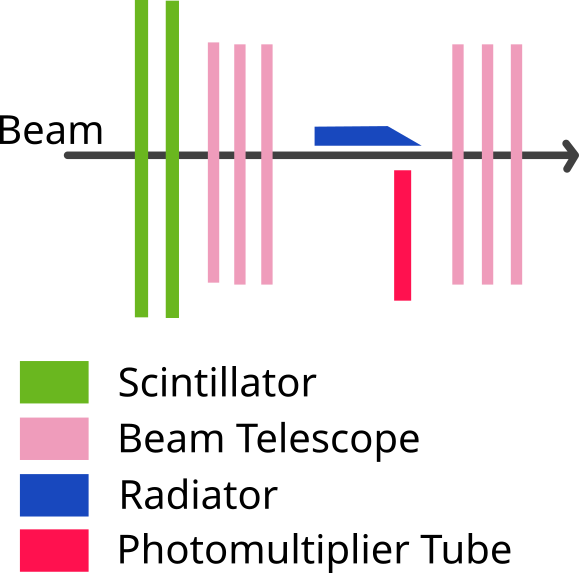} 
 \caption{Sketch of the experimental setup at the DESY II Test beam facility.}
  \label{fig:ExpSetup}
\end{figure}

\begin{figure}
\centering
 \includegraphics[width=0.5\textwidth]{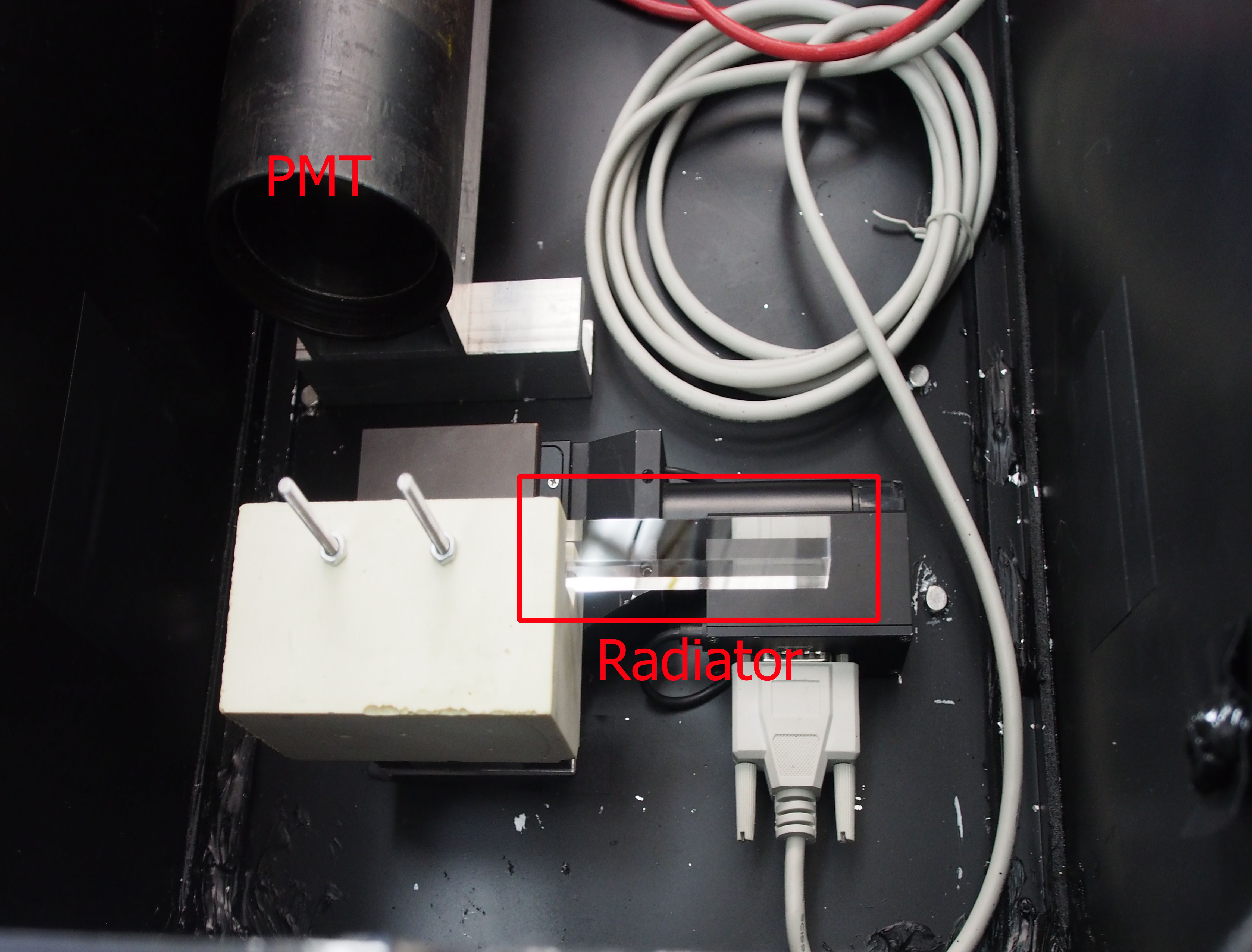} 
 \caption{Picture of the radiator mounted on a movable stage next to the PMT inside a black painted aluminum box.}
  \label{fig:Radiator setup}
\end{figure}

The experimental setup (see Fig.~\ref{fig:ExpSetup}) comprises a beam telescope consisting of six silicon pixel detectors~\cite{Telescope2016} that are permanently installed at DESY, a photomultiplier tube (PMT) and a fused silica radiator. The beam telescope features a high resolution in the order of a few micrometers, and low material budget, which enables the reconstruction of particle tracks at the given momentum range and thus an estimation on the relative particle distance to the radiator. It is used in a configuration (see Table \ref{tab:DetectorPositions}) with three detector planes each before and behind the radiator.
In addition, a pair of scintillators is utilised as input to the trigger system. The centerpiece of the experiment, the radiator and the PMT, can be seen in Figure~\ref{fig:Radiator setup}.  The PMT (ET enterprises 9813QKB) was operated at $\SI{1650}{\V}$ for all experiments.\\

The radiator is positioned partially inside the beam, such that the center of the beam spot is located at the edge of the radiator. Thus, the majority of the particles passes in close proximity of the radiator. Inevitably, a significant fraction of particles traverses the radiator, leading to emission of non-diffraction Cherenkov radiation.
Using the track information obtained via the beam telescope, events of this type can be identified. To reduce contamination from ambient light, PMT and radiator were placed in an aluminum box, painted black on the inside (see Fig.~\ref{fig:Radiator setup}).
The box was placed on linear motion stages for an alignment transverse to the beam, while the radiator itself was mounted on a rotation stage for an angular alignment parallel to the beam.
Beam windows covered with black tape were added to reduce the material budget while maintaining the blocking of ambient light.\\

The PMT can be further equipped with polarization filters in order to study radiation polarization. ChDR is polarized as it arises from fields of charged particles inducing dynamic polarization currents at the air-radiator interface~\cite{Shevelev2014}. The angular distribution is determined by the spatial arrangement of particle beam and radiator~\cite{Shevelev2014}. 
\subsection{Radiator}
\begin{figure}
\centering
 \includegraphics[width=0.5\textwidth]{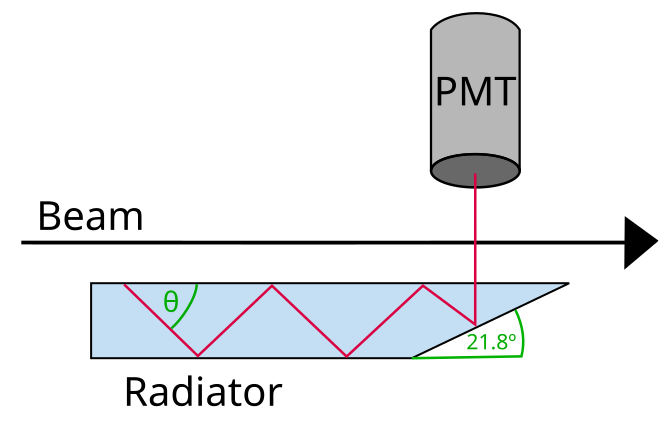} 
 \caption{Sketch of ChDR emission in a long dielectric radiator based on a design previously described for non-invasive beam diagnostics~\cite{Alves2019}.}
  \label{fig:SketchRadiator}
\end{figure}

Right-angled trapezoid prism radiators made of high purity fused silica (SiO$_{2}$) were obtained from Heraeus~\cite{Heraeus} and CERN. The dimensions were $\SI{15}{\cm} \times \SI{1.5}{\cm} \times \SI{1}{\cm}$ and $\SI{5}{\cm} \times \SI{1}{\cm} \times \SI{0.5}{\cm}$, respectively.
The unique prism geometry of the radiators allows for ChDR generated over the entire length of the radiator to reach the PMT. Due to the a significant fraction of the light undergoing internal reflection, it will reach the wedge shaped end of the radiator (see Fig~\ref{fig:SketchRadiator}).
A reflective coating at the $\SI{21.8}{\degree}$ angled surface has been applied to enforce the radiation exiting the radiator perpendicularly to the opposite surface. To determine the QDC signal baseline, a small piece of aluminum foil, blocking the exiting light from entering the PMT, was temporarily applied over this area.

\subsection{Triggering \& Data Acquisition}
Two assemblies of scintillators, light guide and PMT, were used for triggering purposes.
These scintillators were powered and their signals interpreted by the Trigger Logic Unit (TLU)~\cite{TLU2019}, which is used for a coincidence detection on discriminated input signals with a programmable threshold.
The TLU in turn forms a particle trigger signal and performs a trigger-busy-handshake with all detectors, inhibiting any further trigger signals from being distributed while any detector is indicating a busy signal. In consequence of a trigger, the telescope data is recorded, storing the data as well as the trigger numbers to the disk. 
The PMT signal is digitized via a $\SI{12}{\bit}$ charge to digital converter (QDC) of the type CAEN V965~\cite{CAEN}.
For this, an integration window is created through a pulse generator, initiated by the trigger signal and with a width that was empirically determined to cover the full duration of all PMT pulses. The QDC raises a busy signal while the integration is in process. The data acquisition is controlled via the EUDAQ2 data acquisition framework~\cite{EUDAQ}.
This software enables the initialization, configuration and control of the telescope, the QDC, the scintillators, the TLU, and the motion and rotation stages via dedicated configuration files.
It furthermore features so-called producers, which have the task of writing the data to disk.

\begin{figure}
\centering
 \includegraphics[width=0.5\textwidth]{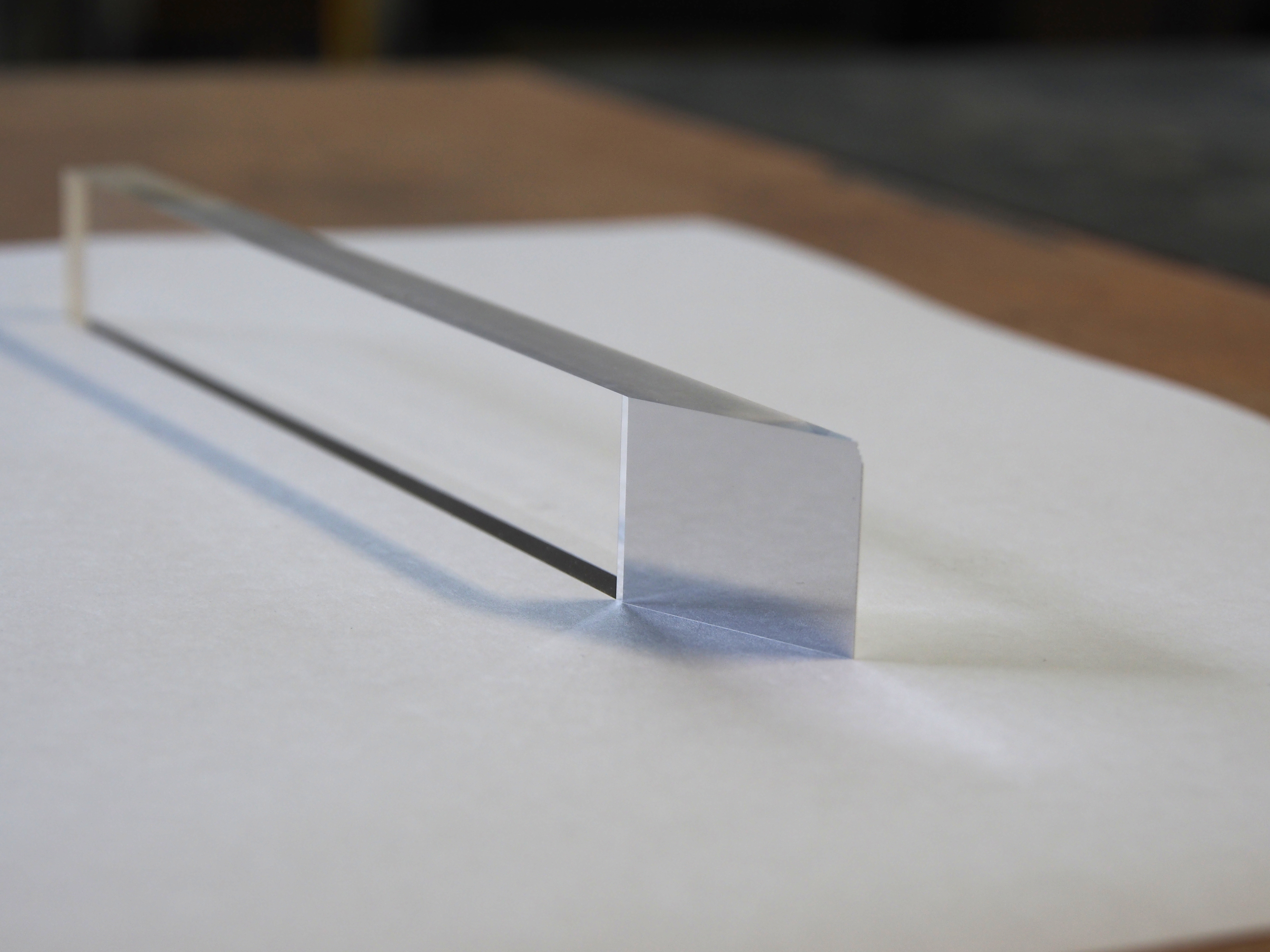} 
 \caption{Picture of a fused silica radiator used in the experiments with the geometry described in Figure \ref{fig:SketchRadiator}.}
  \label{fig:Radiator}
\end{figure}

\subsection{Data Analysis}
\begin{figure}
\centering
 \includegraphics[width=
0.95\textwidth]{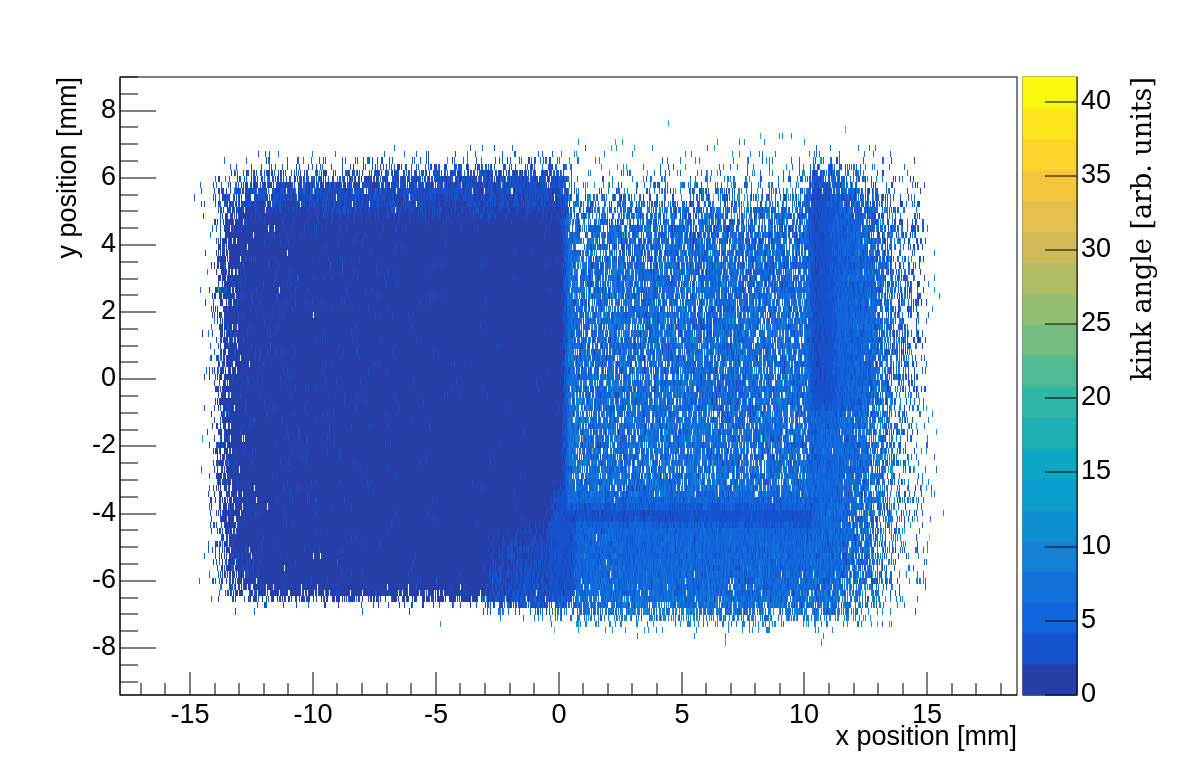} 
 \caption{X-Y projection material budget image of the beam profile kink angle. Both the radiator and the mounting structure are clearly distinguishable from the background.}
  \label{fig:MBI79}
\end{figure}

The data was analysed using ROOT~\cite{ROOT} and PyRoot in Jupyter Notebook \cite{Python}. The particle hits were clustered and tracked using the corryvreckan library~\cite{Dannheim2021}. The position of the particles at the level of the radiator was calculated by reconstructing the pathway of the particle with straight lines fitted through the clusters before and after the radiator. The particle tracks were then used to calculate the impact parameter (distance of the particle from the radiator surface). Positive values indicate a position inside the radiator, negative values indicate a position outside of the radiator. The position of the radiator edge was determined using Material Budget Image (MBI) (see Fig.~\ref{fig:MBI79}), representing a two-dimensional mapping of the amount of material traversed by relativistic charged particles~\cite{Jansen2018}. This was achieved by fitting a straight line through the points on the MBI, where the value of the kink angle reached the midpoint between the minimum and the maximum value.\\

\begin{figure}
\centering
 \includegraphics[width=
0.95\textwidth]{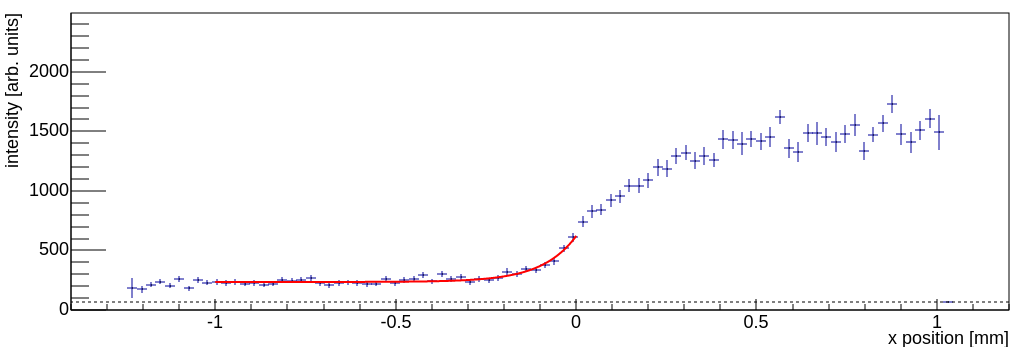} 
 \caption{Exponential fit of the photon emission as a function of impact parameter using $\SI{3}{\GeV\per\c}$ in the intervall $[-1,0]$.}
  \label{fig:Electronfit}
\end{figure}

Next, the pulse amplitude measured from the PMT was plotted over impact parameter. Events with impact parameters greater than $\SI{1}{\mm}$ or smaller than $\SI{-1.2}{\mm}$ were excluded from the analysis as they were considered to be too far away from the air-radiator interface. The graph obtained was then modelled by an exponential function~\cite{Kieffer2018} of type $a+b\cdot\mathrm{e}^{cx}$ with an interval of length $\SI{1}{\mm}$, ending at the previously determined radiator edge (see Fig \ref{fig:Electronfit}). The integral of the exponential function in this interval was used as a single number to compare the total ChDR emission under different experimental conditions. Finally, the arbitrary QDC output unit was converted to photons using a calibration run with a controlled light emission (see Section~\ref{sec:QDC output}). We then performed a linear regression on the photon generation as a function of particle momentum in $R$ (see Fig \ref{fig:Integrals}). The obtained regression line coefficients were compared using analysis of covariance (ANCOVA). The $\SI{6}{\GeV\per\c}$ electron measurement was excluded from the analysis as after tracking there was an insufficient number of data points with accurate tracks (see Fig \ref{fig:Electrons}). This is due to scattering being observed more frequently at high particle momenta. The experiment with $\SI{6}{\GeV\per\c}$ positrons however included enough data.\\

\begin{figure}
\centering
 \includegraphics[width=0.95\textwidth]{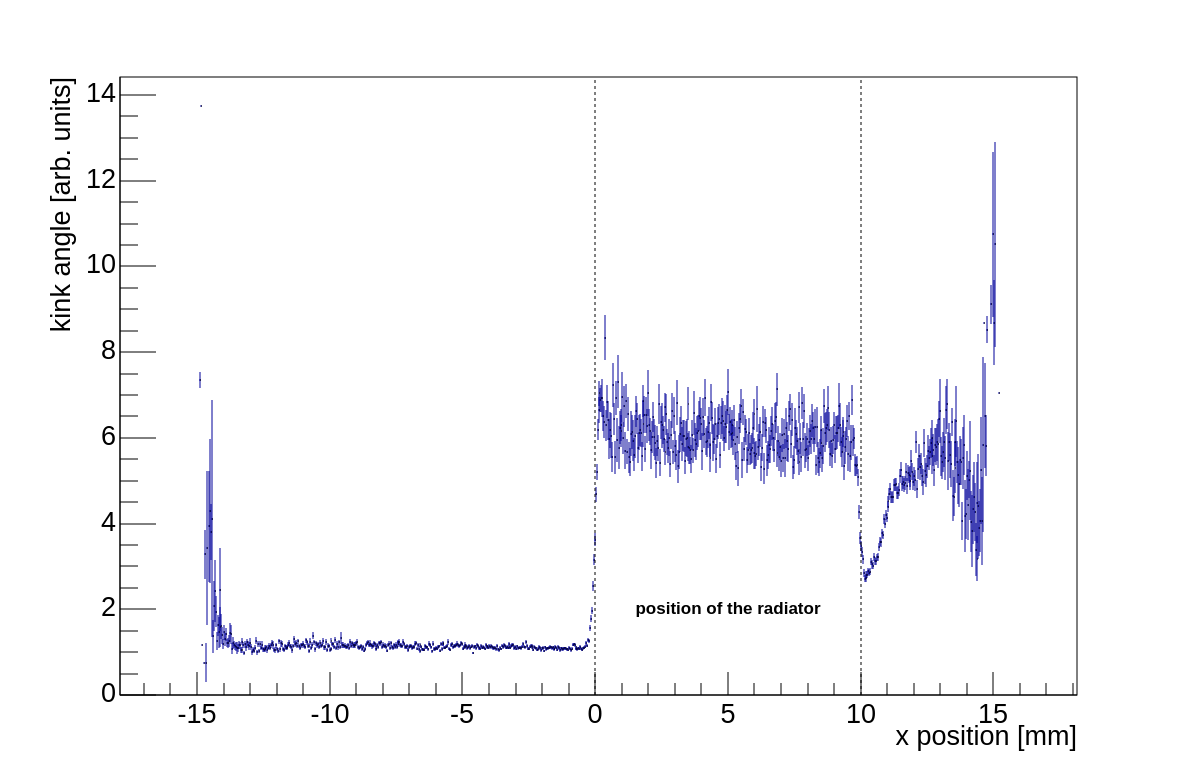} 
 \caption{Centered X projection of the MBI with the determined boundary of the radiator. Non-diffraction Cherenkov radiation is produced inside the radiator that is located between $x=\SI{0}{mm}$ and $x=\SI{10}{\mm}$. For $x$ greater than $\SI{10}{\mm}$, non-diffraction Cherenkov radiation is produced from an interaction between the beam and the mounting stage. For $x$ smaller than $\SI{0}{\mm}$, only ChDR is observed.}
  \label{fig:Projection79}
\end{figure}
\subsection{QDC output calibration}
\label{sec:QDC output}
A pulsed green LED next to the PMT was used to calibrate the QDC output. When the LED is turned on, if it is in a state where few photons reach the PMT, the amount of photons impinging the PMT can be assumed to take a Poisson distribution. By controlling the distance of a LED to the PMT, the voltage of the pulse applied to the LED and the pulse duration, the amount of photons reaching the PMT can be controlled.
In this setup, the data acquisition is triggered by every pulse provided from the LED. Less than 1 in 10 events were observed to have a signal pulse from the PMT, suggesting the Poisson mean is less than $0.1$. Under these circumstances, the probability of having more than one photon is vanishingly small, therefore, the few pulses from the PMT must be from single photon events. By fitting a superposition of two Gaussian functions to the peak in the signal that corresponds to single photons and the bias current pedestal (see Fig.~\ref{fig:QDCPhoton}), an offset value arising from the measurement itself, an estimate of the unit conversion factor can be obtained. The conversion factor measured from the pedestal peak was found to be $\SI{7.2 \pm 3.8}{\QDCunits \per \photon}$. The uncertainty of the conversion factor was calculated using Gauss' law of error propagation and the width of each summand.

\section{Results}
\begin{figure}
\centering
 \includegraphics[width=\textwidth]{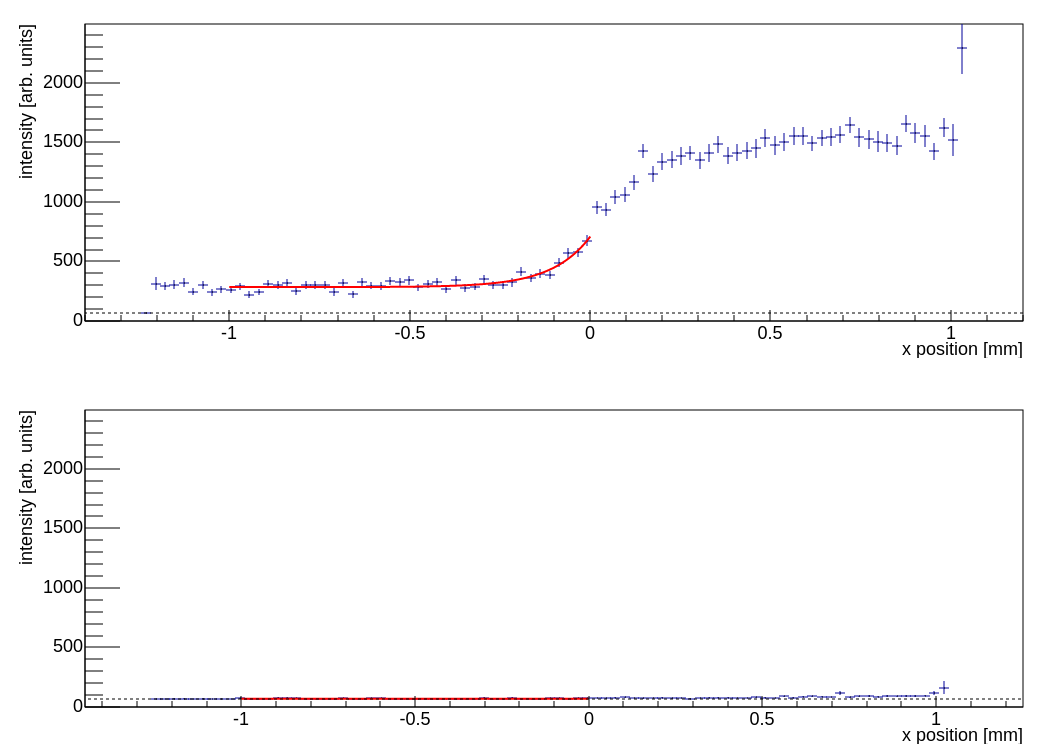} 
 \caption{Exponential fit of the photon emission as a function of impact parameter using $\SI{5}{\GeV\per\c}$ electrons with (bottom) and without (top) aluminum foil on the radiator. Blocking the photon exit point on the radiator reduces both non-diffraction Cherenkov Radiation and ChDR detection to negligible levels.}
  \label{fig:ElectronFoil}
\end{figure}

\begin{figure}
\centering
 \includegraphics[width=0.95\textwidth]{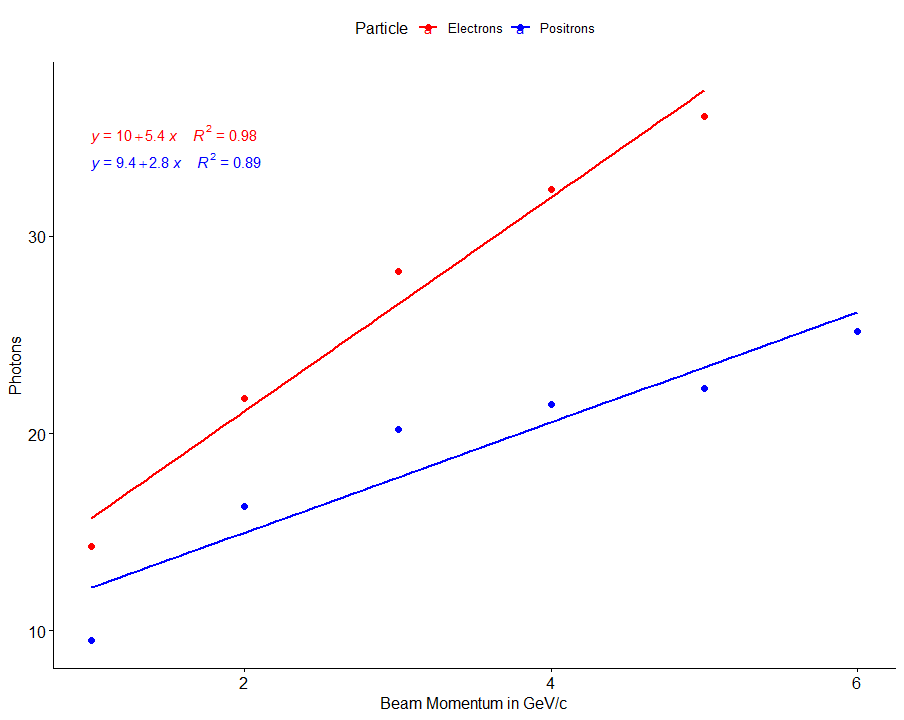}
 \caption{Values of the integral of the exponential fits as a function of beam momentum.}
  \label{fig:Integrals}
\end{figure}

Comparing the experiments with and without aluminum foil (see Fig.~\ref{fig:ElectronFoil}) suggests a significant increase in photon emission due to Cherenkov Radiation from an interaction between the charged particles and the radiator. Tracking the particles allows accurate discrimination between photons generated from non-diffraction Cherenkov Radiation and Cherenkov Diffraction Radiation. We are therefore confident to have detected ChDR. This is supported by our observation of a linear increase in light emission between $\SI{1}{\GeV\per\c}$ and $\SI{6}{\GeV\per\c}$ for positrons and $\SI{1}{\GeV\per\c}$ and $\SI{5}{\GeV\per\c}$ for electrons when comparing the values of the integral of the exponential fit (see Fig~\ref{fig:Integrals}).\\

Figure~\ref{fig:Integrals} also suggests electron ChDR emission is more dependent on particle momentum than that of positrons, resulting in a greater slope. To validate this, we performed an ANCOVA test on the regression lines. We found a significant difference between electrons and positrons after adjustment for particle momentum ($p=0.000862$).  This is supported by visual differences in the photon emission rates as a function of $x$ position (see Fig~\ref{fig:Electrons} \& \ref{fig:Positrons}).\\

To further characterise the radiation, we measured the photon generation for various orientations of a polarization filter placed over the PMT. The results from this data indicate that ChDR generated in our setup has a higher horizontal than vertical polarization component both at $\SI{3}{\GeV\per\c}$ and $\SI{5}{\GeV\per\c}$ (see Table~\ref{tab:polarisation}). Higher emissions of vertically polarized photons for a radiator placed above the beam have been reported in the literature~\cite{Kieffer2020}. As the radiator in our experiment was placed on the side of the beam, our results agree with this observation.\\

We also evaluated a short radiator from CERN on top of the Heraeus radiator we used for the main experiments. The dimensions were $\SI{5}{\cm} \times \SI{1}{\cm}\times \SI{0.5}{\cm}$ and  $\SI{15}{\cm} \times \SI{1.5}{\cm} \times \SI{1}{\cm}$, respectively. Previous experiments suggested a linear increase in light emission for longer radiators~\cite{Alves2019}. We found an increase in light emission (see Table~\ref{tab:positronselectrons}) for the larger radiator, but this was higher than the tripling the theory predicts for a radiator with trice the length. The deviation may be due to differences in thickness, width or manufacturing of the radiators.

\begin{table}
\centering
\begin{tabular}[h]{l|c|c}
Polarisation & $\SI{3}{\GeV\per\c}$ & $\SI{5}{\GeV\per\c}$ \\
\hline
Vertical            & 0.39 & 0.40 \\
Vertical +45° ccw   & 0.77 & 0.88 \\
Horizontal          & 3.12 & 3.78 \\
Horizontal +45° ccw & 1.54 & 1.90
\end{tabular}
\caption{ChDR emission in photons for $\SI{3}{\GeV\per\c}$ and $\SI{5}{\GeV\per\c}$ electrons using different orientations of the polarisation filters.}
\label{tab:polarisation}
\end{table}

\begin{table}
\centering
\begin{tabular}[h]{l|c|c}
Beam & Small radiator & Large radiator \\
\hline
2 GeV $e^-$ & 1.99 & 21.799 \\
3 GeV $e^-$ & 2.35 & 28.19 \\
2 GeV $p^+$ & 2.15 & 16.28 \\
3 GeV $p^+$ & 2.40 & 20.16
\end{tabular}
\caption{ChDR emission in number of photons for $\SI{2}{\GeV\per\c}$ and $\SI{3}{\GeV\per\c}$ electrons and positrons using two different radiators.}
\label{tab:positronselectrons}
\end{table}

\section{Conclusion}
We show that ChDR emission increases linearly with particle momentum between $\SI{1}{\GeV\per\c}$ and $\SI{5}{\GeV\per\c}$ for both positrons and electrons. Unlike previous experiments on circular colliders, we measured emission by both particle types in the same setup. We report a significantly higher increase in ChDR emission rates as a function of particle momentum for electrons compared to positrons. To our knowledge, differences in emission rates of electrons and positrons have not been reported for ChDR or non-diffraction Cherenkov Radiation. Further experiments to investigate this possible difference are needed. Our results also indicate that ChDR may be useful for monitoring the momenta of particle beams, as the light emissions are a linear function of the particle momentum for both positrons and electrons.

\section{Acknowledgements}
The students among the authors would like to thank their teachers Mr. Seidemann and Mr. Irmer for taking them to CERN and BESSY II and sharing their passion for physics with them. They would also like to thank Sarah Aretz and Margherita Boselli for organizing the competition as well as all volunteers from DESY and CERN for supporting the data analysis. 
The students are thankful for financial support by the CERN and Society Foundation, the Wilhelm and Else Heraeus Foundation, the Arconic Foundation, AMGEN, and the Ernest Solvay Fund, managed by the King Baudouin Foundation. They would also like to express their gratitude towards CERN and DESY for organising BL4S. Receival of radiators from Thibaut Lefèvre of CERN and Heraeus Group is gratefully acknowledged.
The measurements leading to these results have been performed at the Test Beam Facility at DESY Hamburg (Germany), a member of the Helmholtz Association (HGF).
\section{Declaration of Competing Interest}
The authors declare that they have no known competing financial interests or personal relationships that could have appeared to influence the work reported in this paper.
\section{Data availability}
Data will be made available on request.
\newpage

\appendix
\section{}
\label{sec:appendix}

\begin{table}[H]
\centering
\begin{tabular}[h]{l|c}
Detector &  Z-Position\\
\hline
Detector 1  & $\SI{0.0}{\cm}$ \\
Detector 2  & $\SI{10.2}{\cm}$  \\
Detector 3  & $\SI{20.5}{\cm}$ \\
Detector 4  & $\SI{59.0}{\cm}$ \\
Detector 5  & $\SI{71.7}{\cm}$ \\
Detector 6  & $\SI{85.5}{\cm}$
\end{tabular}
\caption{Z-Positions of the 6 beam telescope silicon pixel detectors.}
\label{tab:DetectorPositions}
\end{table}

\begin{figure}[H]
\centering
\includegraphics[width=\textwidth]{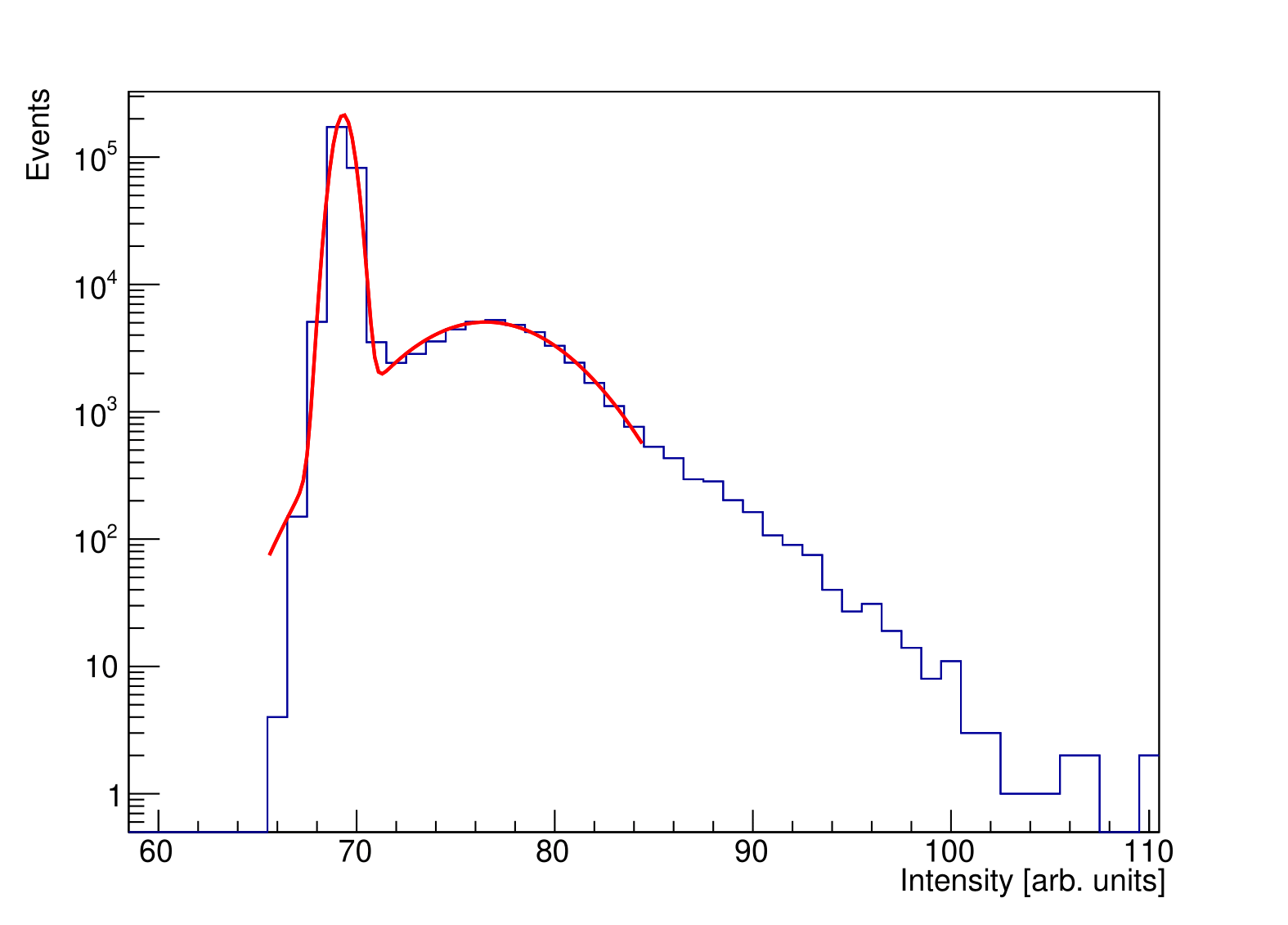} 
\caption{Measured intensity distribution from a flashing single photon LED (blue) fitted with a superposition of two Gaussian functions (red).}
 \label{fig:QDCPhoton}
\end{figure}
\begin{figure}[H]
\centering
 \includegraphics[width=\textwidth]{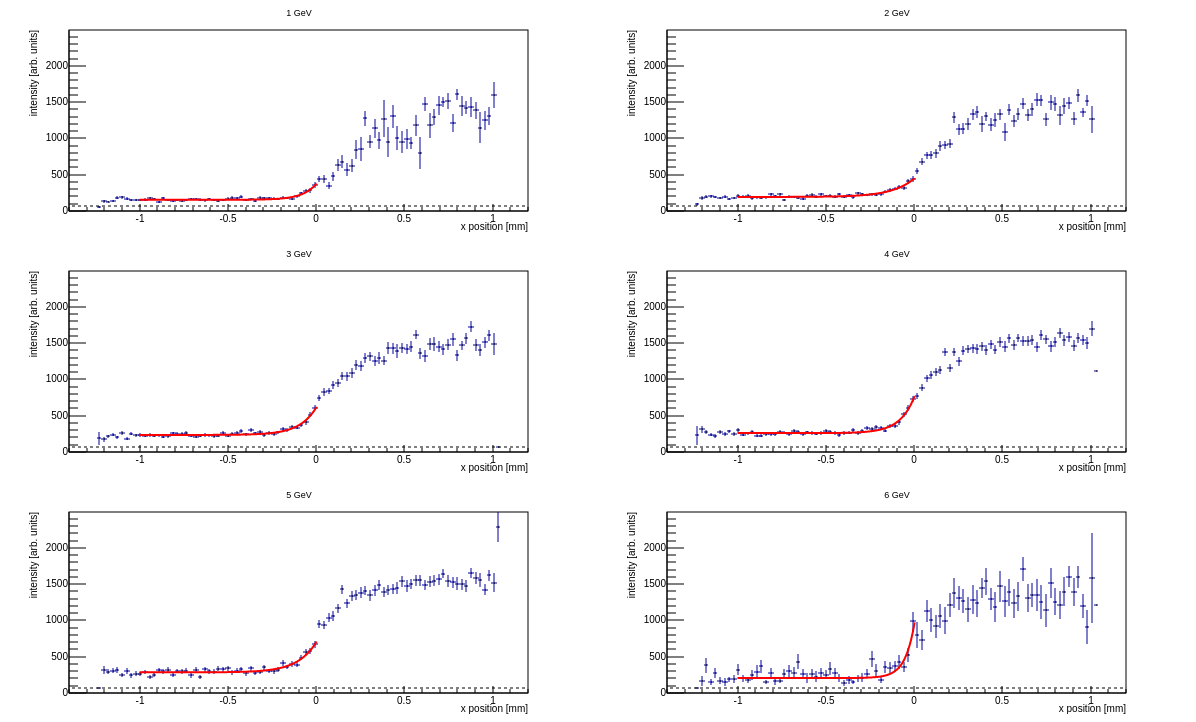} 
 \caption{Exponential fit of the photon emission by electrons as a function of impact parameter for $\SI{1}{\GeV\per\c}$ to $\SI{6}{\GeV\per\c}$.}
  \label{fig:Electrons}
\end{figure}

\begin{figure}[H]
\centering
 \includegraphics[width=\textwidth]{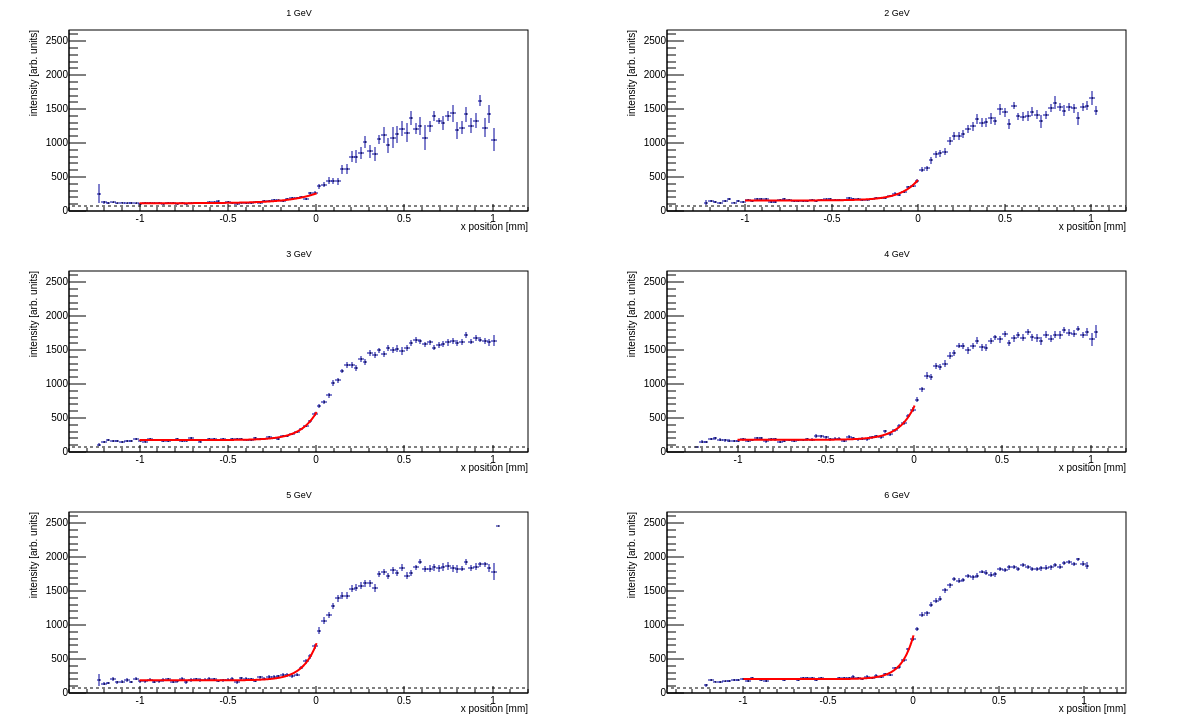} 
 \caption{Exponential fit of the photon emission by positrons as a function of impact parameter for $\SI{1}{\GeV\per\c}$ to $\SI{6}{\GeV\per\c}$.}
  \label{fig:Positrons}
\end{figure}

\newpage

\bibliographystyle{elsarticle-num} 
\bibliography{cas-refs}





\end{document}